\documentclass[aps,pre,superscriptaddress,showpacs]{revtex4}

\usepackage{makeidx}
\usepackage{graphicx}
\usepackage{amssymb}
\usepackage{amsbsy}
\usepackage[dvips]{color}

\newcommand{\beq}{\begin{equation}}
\newcommand{\eeq}{\end{equation}}
\newcommand{\be}[1]{\begin{equation}\label{#1}}
\newcommand{\ee}{\end{equation}}
\newcommand{\continue}{\nonumber \\ }
\newcommand{\bea}{\begin{eqnarray}}

\newcommand{\eea}{\end{eqnarray}}

\newcommand{\Tr}{\mbox{Tr}}

\renewcommand{\det}{\mbox{\rm det}}

\newcommand{\tr}{{\rm tr}\, }

\newcommand{\FIG}[4]{\begin{figure}
                      \begin{minipage}[l]{0.90\textwidth}
                      \noindent{#1}

                      \caption[#2]{#3}

                      \label{#4}
                      \end{minipage}
                      \end{figure} }

\newcommand{\rf}     [1] {~\cite{#1}}

\newcommand{\refRef} [1] {Ref.~\cite{#1}}

\newcommand{\refeq}  [1] {(\ref{#1})}
\newcommand{\refeqs} [2]{(\ref{#1}--\ref{#2})}

\newcommand{\reffigs} [2] {Figs.~\ref{#1} and~\ref{#2}}
\newcommand{\refFig} [1] {Fig.~\ref{#1}}

\newcommand{\refTab} [1] {Tab.~\ref{#1}}

\newcommand{\refSect}[1] {Sect.~\ref{#1}}

\newcommand{\gf}{\mathbf{G}}

\newcommand{\proj}{\mathbf{e}}
\newcommand{\normal}{\mathbf{n}}

\begin{document}


\title{Counting Function for a Sphere of Anisotropic Quartz}


\author{Niels S\o ndergaard}
\email[]{Niels.Sondergaard@matfys.lth.se}
\author{Thomas Guhr}
\email[]{Thomas.Guhr@matfys.lth.se}
\affiliation{Matematisk Fysik, LTH, Lunds Universitet, Sweden}

\author{Mark Oxborrow}
\email[]{Mark.Oxborrow@npl.co.uk}
\affiliation{National Physics Laboratory, Teddington, UK}

\author{Kristian Schaadt}
\email[]{Schaadt@nbi.dk}
\author{Clive Ellegaard}
\email[]{Ellegaard@nbi.dk}
\affiliation{Niels Bohr Institute, Copenhagen, Denmark}


\date{\today}

\begin{abstract}
We calculate the leading Weyl term of the counting function for a
mono-crystalline quartz sphere. In contrast to other studies of
counting functions, the anisotropy of quartz is a crucial element in
our investigation.  Hence, we do not obtain a simple analytical form,
but we carry out a numerical evaluation. To this end we employ the
Radon transform representation of the Green's function. We compare our
result to a previously measured unique data set of several tens of
thousands of resonances.
\end{abstract}

\pacs{05.45.Mt, 43.20.+g, 43.40.+s}

\maketitle


\section{Introduction}

In solid state physics \rf{AM}, quantum
chaos\rf{Predrag,gutbook,brack,Stock} and other applications a smooth
approximation to the level density or, equivalently, to its integral
referred as the counting function (or staircase function) is often
needed.  Typically such an approximation is obtained by semiclassical
methods\rf{BaltHil}.  Alternatively, related techniques stemming from
geometrical quantization and index theorems\rf{GeoQu} are occasionally
employed. In the context of quantum chaos, this problem of counting
the number of states has been studied thoroughly for finite systems,
in particular, billiards. The latter are realized in microwave
resonators. Quantum chaos methods were successfully applied to
elastodynamical
systems\rf{Weaver,alu,quartz,para,waves,couch,disc,HB}. The
semiclassical limit corresponds to a ray limit in which the wave
lengths are much smaller than characteristic length scales of the
resonating device. However, anisotropy enters as an important feature
of these systems which is not present in scalar Schr\"odinger quantum
mechanics. The problem of anisotropy also shows up in the calculation
of the heat capacity in solid state physics which directly relates to
the counting function.

In this contribution, we study a mono-crystalline quartz sphere.  We
calculate the first approximation to the counting function which is
referred to as the Weyl term in semiclassics.  We combine analytical
with numerical computations.  Furthermore, we have a unique data set
at our disposal which was measured previously with extraordinary
resolution~\cite{MCK}. It comprises tens of thousands of resonances,
enabling a comparison with our theoretical findings.

Even though the particular shape of a sphere was chosen the system
will not be rotationally invariant due to the underlying medium,
$\alpha-$quartz, which is anisotropic.  This anisotropy leads to
non-separabillity of the wave equation. This makes the problem so
complex that there is no way of calculating spectra than by brute
force numerical methods.

A major motivation for the present investigation is the renewed
interest in elastomechanical problems which are increasingly important
for micro--electro--mecha\-ni\-cal systems (MEMS).  In particular, a
better understanding of spectral properties is expected to improve the
control of the resonance and scattering pole structure of MEMS. This
in turn would allow designing better filters and
actuators\rf{MEMS}. Thus deeper insight into the role of anisotropy is
called for.

The article is organized as follows.  In \refSect{elastoSect}, we
compile general features of elastodynamics as needed in the present
context.  We work out the Green function in \refSect{gfSect}.  The
data accumulation is briefly reviewed in \refSect{data}, we compare to
the experimental findings in \refSect{weylSect}.  We summarize and
conclude in \refSect{summarySect}.

\section{Elastodynamics}
\label{elastoSect}

In Sect.~\ref{sec21}, we sketch some properties of elastic waves in
anisotropic materials. We discuss the dispersion relation and the
slowness surfaces in Sect.~\ref{sec22}. The Green's function is worked
out in Sect.~\ref{sec23}.  The Einstein summation convention is
assumed throughout.

\subsection{The elastic wave equation} 
\label{sec21}

In linear elastodynamics in the time domain, the wave equation reads
\rf{lAndl,auld}
\be{waveTime}
c_{ijkl} \, \frac{\partial^2 u_l}{\partial x_j \partial x_k} = \rho
  \frac{\partial^2 u_i}{\partial t^2} \, ,
\ee
where the displacement field $u_i$ describes the increment of a
fictitious point particle at its position $(x_1,x_2,x_3)$. The
quantities $c_{ijkl}$ form the elasticity tensor and $\rho$ is the
mass density. For our system, we may assume homogenity of the
material, that is, the mass density and the elasticity tensor are
constants.  In an isotropic material, the elasticity tensor has only
two independent entries which yield, together with the mass density,
the longitudinal velocity for pressure waves and the transverse
velocity for shear waves.  In our case of anisotropic quartz, the
velocity of an elastic wave also depends on the direction in the
crystal. This means that the elasticity tensor has six independent
entries.  The elasticity tensor always fulfills the following symmetry
conditions
\beq
c_{ijkl}=c_{ijlk}=c_{klij} \, .
\eeq
The wave equation~(\ref{waveTime}) is valid in the regime where the
linear relation
\be{Hooke}
\sigma_{ij} = c_{ijkl} u_{kl}
\ee
holds between the stress tensor $\sigma_{ij}$ and the strain tensor
\be{strain}
u_{kl} = \frac{1}{2} \left(\frac{\partial u_k}{\partial x_l} +
             \frac{\partial u_l}{\partial x_k}\right) \, . 
\ee
Equation~(\ref{Hooke}) is the most general form of Hooke's law. The
wave equation in the frequency domain which corresponds to
Eq.~(\ref{waveTime}) reads
\be{waveFreq}
c_{ijkl} \, \frac{\partial^2 u_l}{\partial x_j \partial x_k} + \rho
 \, \omega^2 \,  u_i= 0 \, ,
\ee
where the angular velocity is denoted $\omega$.  

We are interested in the elastic vibrations of a finite objects, the
elastic resonator, which is confined by a surface.  Matching the
situation in the experiment we assume free boundary conditions, that
is, there is no normal stress on the surface. This is equivalent to
the condition
\be{tracD}
0 = c_{ijkl} \, u_{k l} \, n_j
\ee 
with the strain~(\ref{strain}) and normal vector $n_j$. The boundary
condition~(\ref{tracD}) is of Neumann type and makes it possible to
mode convert at the boundary. In an isotropic material, a longitudinal
(transverse) wave hitting the boundary is reflected and generates,
apart from some special situations, a second wave which is transverse
(longitudinal). These waves have different velocities and leave the
point of incident under different angles. This formally corresponds to
quantum mechanics for a free spin-one particle whose effective mass
depends on the spin degrees of freedom.  The particle is confined in
an enclosure, corresponding to the resonator. Not surprisingly, mode
conversion shows a much richer phenomenology in an anisotropic
material.

\subsection{Dispersion relation and slowness}
\label{sec22}

We consider a plane wave with wave vector $k_j$, angular velocity
$\omega$ and polarization vector $p_l$,
\be{planewave}
u_l^{({\rm pw})}(x_1,x_2,x_3) = p_l \exp\left(ik_jx_j-i\omega t\right) \ .
\ee 
Inserting this into the wave equation~\refeq{waveTime} yields the
dispersion relations 
\be{dispRelK}
c_{ijkl} k_j k_k \, p_l = \rho \omega^2 \delta_{il} \,  p_l  \, .
\ee
(Here $k_j$ and $k_k$ are the $j$--th and th $k$--th component,
respectively, of the wave vector.) The wave numbers and
the angular velocity occur in equal powers. This is exploited by
considering their ratio, the {\it slowness} \rf{musgrave}
\be{slowness}
s_i = k_i/\omega \ . 
\ee
The dispersion relations~(\ref{dispRelK}) then motivate the
introduction of the matrix
\be{slowmat}
S_{il} = - \frac{1}{\rho} \, c_{ijkl} \, s_j  s_k + \delta_{il} \, .
\ee 
Thus, Eq.~(\ref{dispRelK}) can be cast into the form $S_{il}p_l=0$. 
For solutions to exist, one must have 
\be{slowsurf}
\det \, \mathbf{S} = 0 \, .
\ee
This defines the {\it slowness surface}. As each entry in
Eq.~\refeq{slowmat} is quadratic, the full determinantal condition
becomes a polynomial of order six in the variables $s_i$. This
condition gives rise to a surface depicted in
Figs.~\ref{sheetQL},~\ref{sheetQT1} and~\ref{sheetQT2}.
The elastic constants $c_{ijkl}$ for quartz have been used, see
Sec.~\ref{weylSect}. The slowness coordinates are, except for a
constant factor, coordinates in the space dual to the configuration
space of the resonator. The units on the figures are in inverse
velocities corresponding to velocities of the order $5000 \,\mathrm{
m/sec}$.
\FIG{
\rotatebox{0}{\includegraphics[width=8cm]{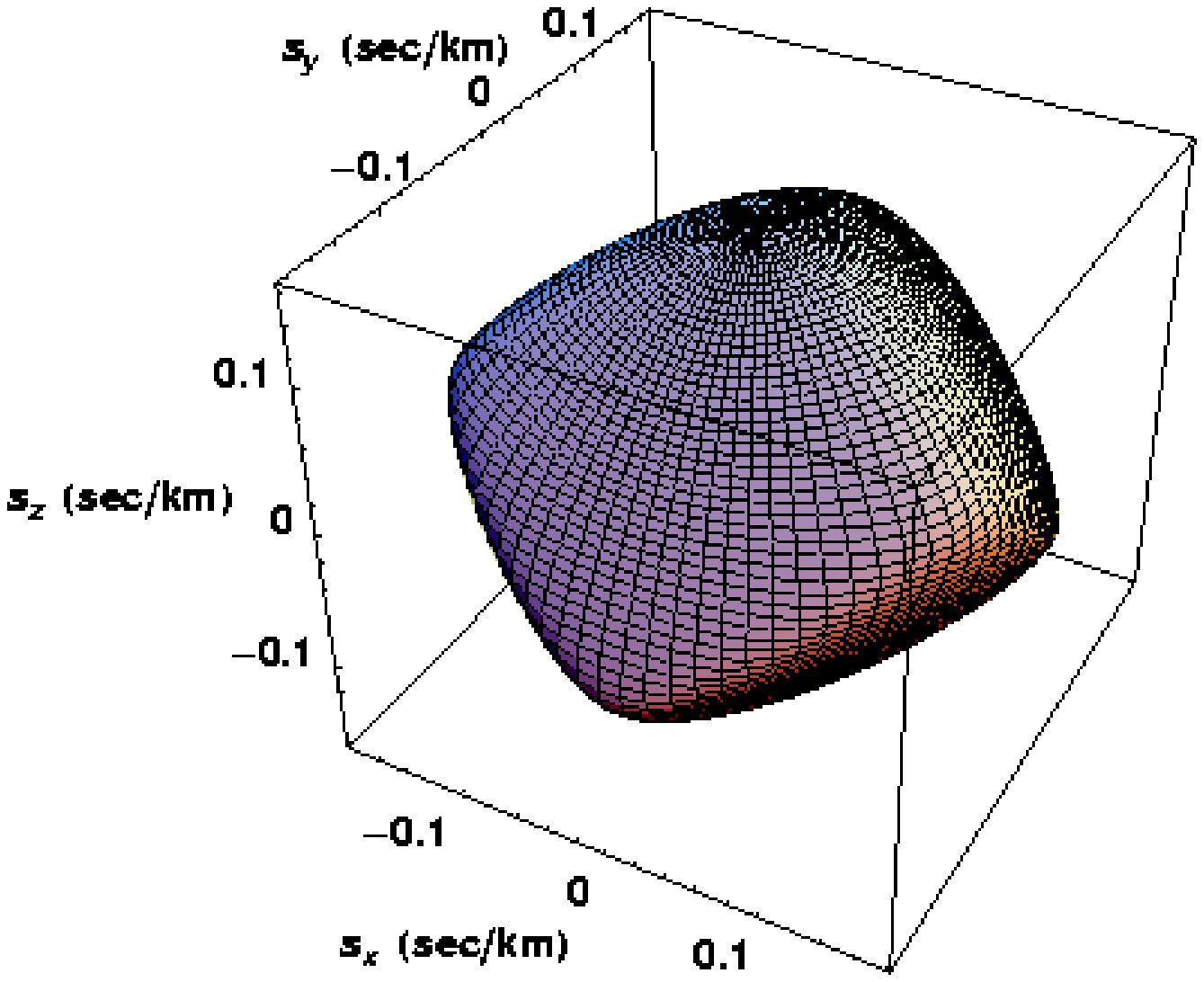}}
}{sheetQL}{ Quasi-Longitudinal slowness.}{sheetQL}
\FIG{
\rotatebox{0}{\includegraphics[width=8cm]{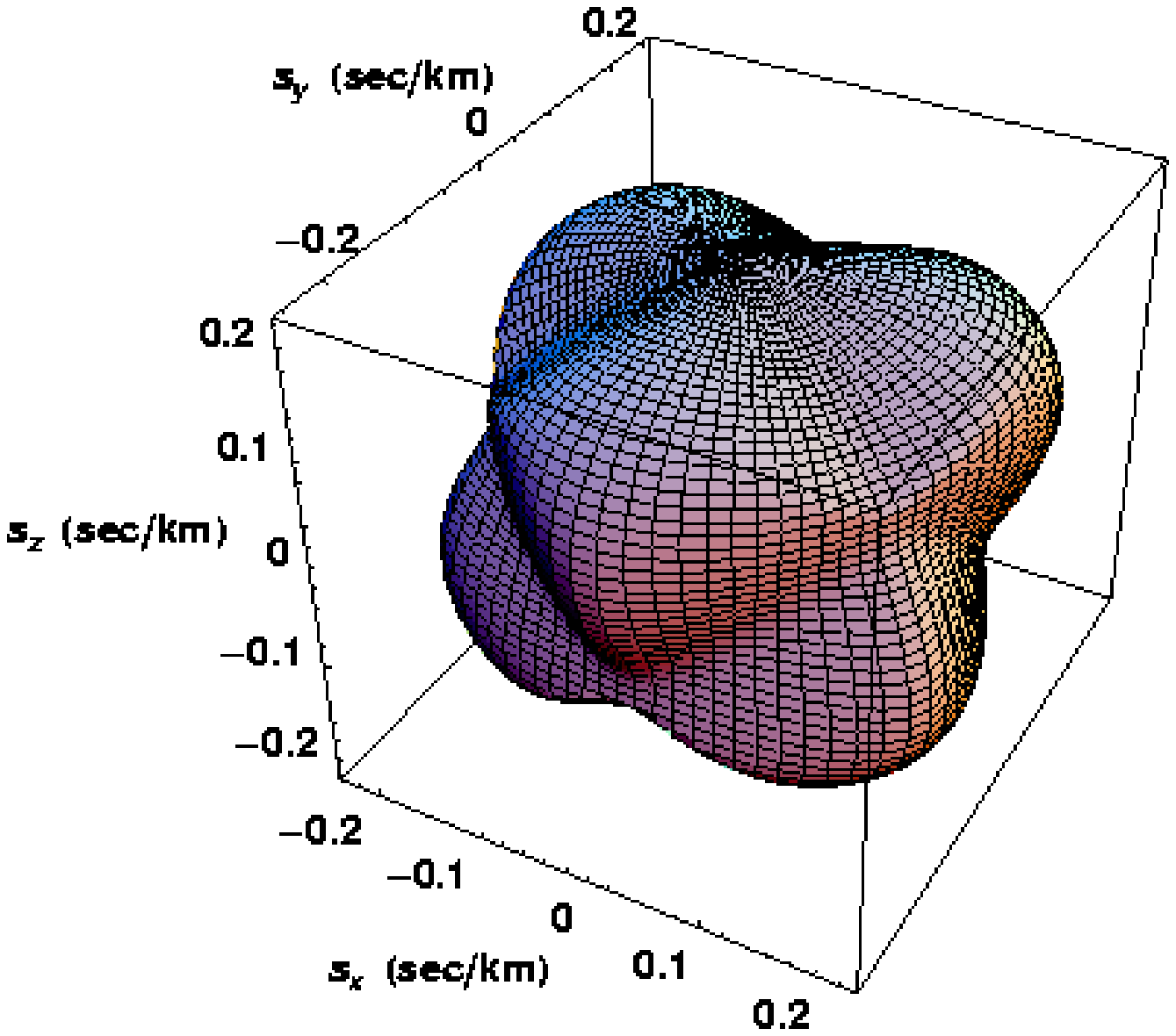}}
}{sheetQT1}{ Quasi-Transverse 1 slowness.}{sheetQT1}
\FIG{
\rotatebox{0}{\includegraphics[width=8cm]{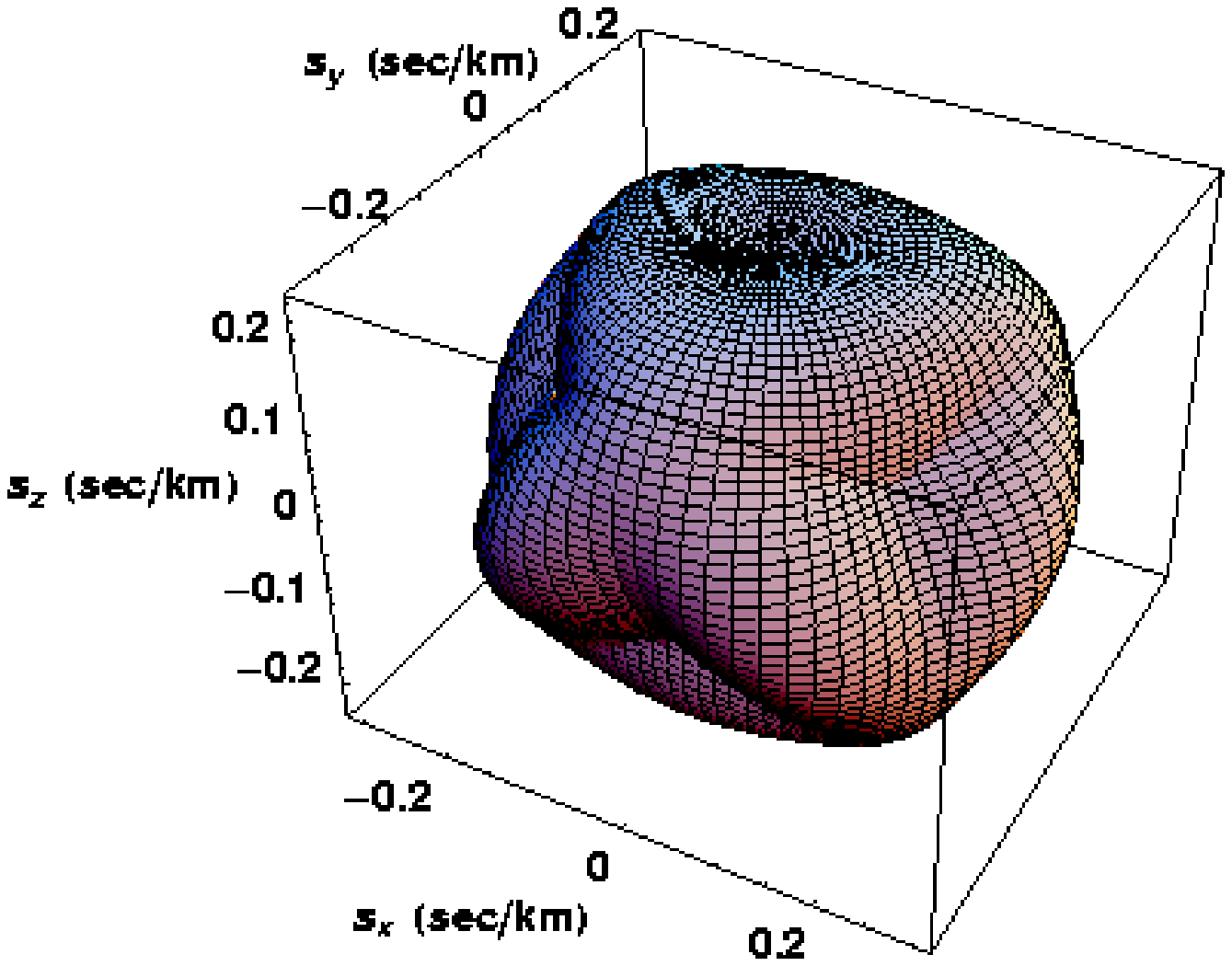}}
}{sheetQT2}{ Quasi-Transverse 2 slowness.}{sheetQT2} 
The surface consists of three sheets, $\Sigma = \cup_{\alpha=1}^3
\Sigma_\alpha$, where only the inner is convex. One clearly sees the
 $D_3$-symmetry of the crystal, that is, after a rotation of $2 \pi/6$
around the symmetry axis a reflection in the plane orthogonal to the
axis transforms the surface into itself. This surface is the remains
of the dispersion surface after the frequency has been scaled
out. Thus a fixed direction corresponds in general to three points on
the slowness surface. The inner sheet is associated to an almost
longitudinal polarization whereas the outer sheets are associated with
the transverse polarization. Hence the terms quasi-longitudinal and
quasi-transverse are used.

As already pointed out, our elastodynamical problem has formally much
in common with quantum mechanics for a particle with spin confined to
an enclosure, which could be viewed as a billiard. Inside the
billiard, the particle moves on straight lines and it mode converts
upon reflections at the boundaries. This is the situation in
configuration space.  Employing slowness, we have mapped the problem
of one spherical billiard for a particle with spin onto the problem of
a particle without spin (in a dual space) simultaneously moving in
three non--spherical billiards, that is, in the spaces confined by the
three slowness surfaces. A trajectory in configuration space consists
of a sequence of straight pieces with a certain modal character. In
slowness space, the particle jumpes between the billiards which are
specific for a given modal character.

\subsection{Group velocity and polarization}
\label{sec23}

The Green's function to be introduced later on is a matrix due to the
vector character of the elastic field. Thus, it is useful to work with
a proper basis in matrix space. Such a basis is formed by the
projectors $\proj_\alpha$ for a given polarization $\alpha$. We shall
consider the case where wave vectors belong to a fixed sheet of
polarization $\alpha$ and discuss the associated group velocity. Below
we denote polarization indices with {\it Greek} letters whereas
geometrical indices are in {\it Roman} letters.

A projector $\proj_\alpha$ is a $3\times 3$ matrix which we introduce,
corresponding to Eqs.~\refeq{dispRelK} and~\refeq{slowmat}, by writing
\beq
\mathbf{k_\alpha \cdot c/\rho \cdot k_\alpha = \omega^2 \, \proj_\alpha} \,, 
\eeq
that is $c_{pqrs}/\rho k_{\alpha,p} k_{\alpha,s}= \omega^2 \,
e_{\alpha, {q r}}$. Each projector corresponds to one of the three
possible polarizations $\alpha$.  The projectors $\proj_\alpha$
satisfy the completeness relation
\beq
\sum_\alpha \proj_\alpha = \mathbf{1}_{3 \times 3} \,
\eeq
and the relation
\beq
 \proj_\alpha \cdot \proj_\beta =  \delta_{\alpha \beta} \, \proj_\alpha \, ,
\eeq
where the dot indicates matrix multiplication.  The dispersion
relation for a fixed polarization becomes
\be{dispRelation}
\mathbf{(k_\alpha \cdot c/\rho \cdot k_\alpha ):\proj_\alpha - \omega^2 = 0 } \, ,
\ee
where the contraction of all indices is denoted by the $:$-sign.

A natural quantity that now enters the discussion is the group
velocity for a given polarization $\alpha$  
\beq
\mathbf{v_{g,\alpha}}  = \frac{\partial \omega}{\partial \mathbf{k}_\alpha}  \, ,
\eeq
describing the change of angular velocity with the wave vector for
that polarization. Varying Eq.~\refeq{dispRelation} with respect to the
wave vectors and the angular velocity
\beq
\mathbf{(\mathrm{d}k_\alpha \cdot c/\rho \cdot k_\alpha ):\proj_\alpha 
              - \mathrm{d}\omega \,  \omega = 0 }
\eeq
and finally scaling out the angular velocity gives
\beq
\mathbf{v_{g,\alpha} = \frac{\partial \omega}{\partial k_\alpha}
              = (c/\rho \cdot s_\alpha) :e_\alpha} \, .
\eeq
Above, also the projectors could have a variation which we have not
taken into account. However, since
\beq
\mathbf{(k_\alpha \cdot c/\rho \cdot k_\alpha ): \mathrm{d}e_\alpha 
    = \omega^2 e_\alpha:\mathrm{d}e_\alpha}  
\eeq
and
\beq
2 \, \mathbf{\proj_\alpha:\mathrm{d}\proj_\alpha \ 
   =  \mathrm{d}( \tr(\proj_\alpha \cdot \proj_\alpha)) 
   =  \mathrm{d}(\tr(\proj_\alpha))} = \mathrm{d}( 1) = 0 \, ,
\eeq
this variation does not contribute. 

{}From Eqs.~\refeq{slowness} and~\refeq{dispRelation} we see that
$\mathbf{s_\alpha}$ and $\mathbf{v_{g,\alpha}}$ are mutually {\it
polar reciprocal} to each other, that is,
\be{polarRec}
1 = \mathbf{(s_\alpha \cdot c/\rho \cdot s_\alpha ):\proj_\alpha  
  = s_\alpha \cdot (c/\rho \cdot s_\alpha) :\proj_\alpha   
  = s_\alpha \cdot v_{g,\alpha}} \, .
\ee
This relation can be used to construct the group velocity and
holds in general for systems which are homogeneous in $k_i$ and
$\omega$. This condition defines the individual sheet, and in
particular the group velocity is normal to this sheet.

However, interesting complications arise as the degeneracy of this
matrix is not constant. Hence there exist directions for which the
phase velocity of the two intersecting sheets is the same.  Note that
this degeneracy occurs in slowness space. Interestingly such {\it mode
conversion points} may also occur purely in $x$-space, e.g. in
Landau-Zener problems or in nuclear/chemical reaction physics. In
general, however, there are three distinct polarizations, and from
Eq.~\refeq{slowmat} these only depend on the direction
$\mathbf{\hat{s}_\alpha} = \mathbf{s_\alpha}/s_\alpha$. Furthermore
for the opposite direction the same polarization is found.
\FIG{
\rotatebox{0}{\includegraphics[height=6cm]{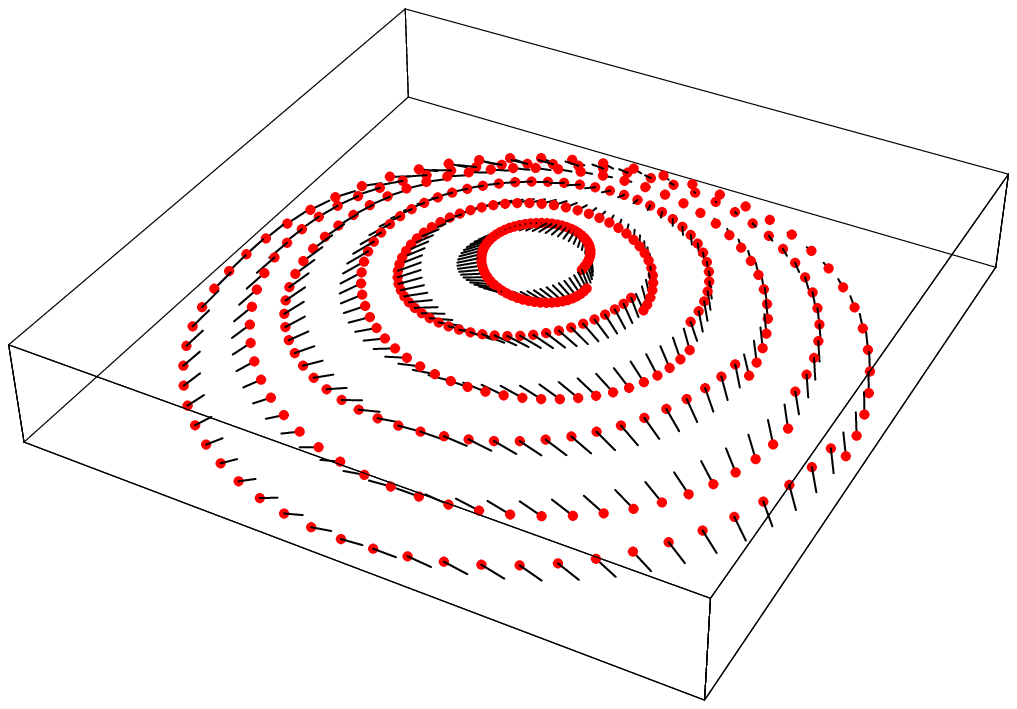}}
}{qsh1}{ Twisting of polarization: Shear 1}{quasiShear1}
\FIG{
\rotatebox{0}{\includegraphics[height=6cm]{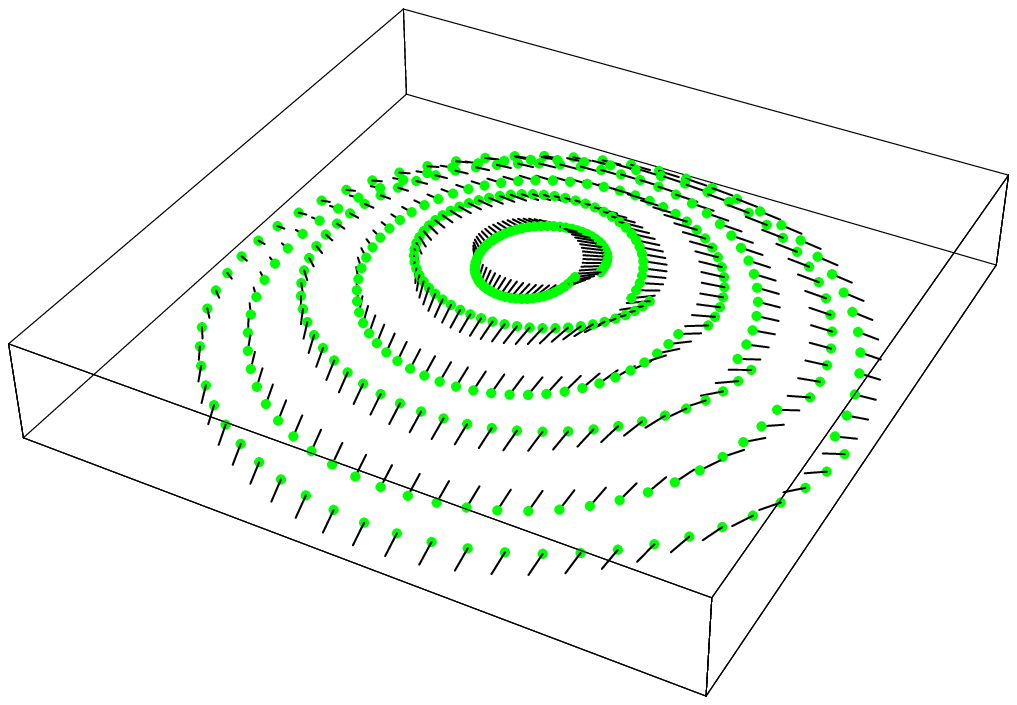}}
}{qsh2}{ Twisting of polarization: Shear 2}{quasiShear2}
\FIG{
\rotatebox{0}{\includegraphics[height=6cm]{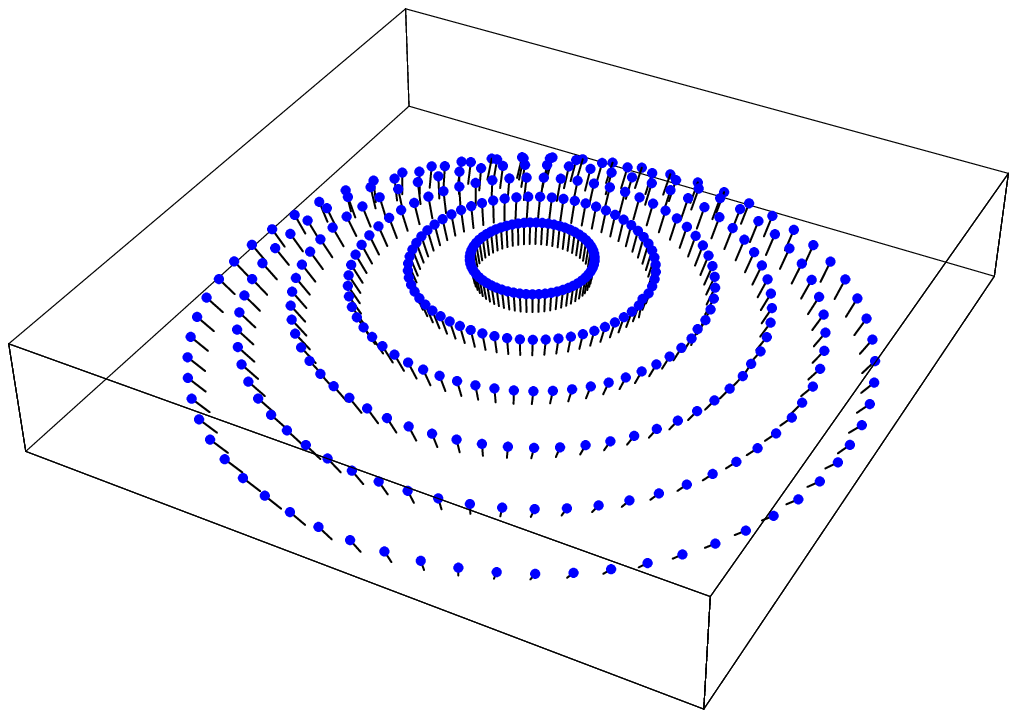}}
}{qpr}{ Polarization: Pressure}{quasiPressure} We have depicted one of
the quasi-transverse polarizations on
\reffigs{quasiShear1}{quasiShear2}. The behavior of the two shear
components are similar with a twisting near the poles. The
quasi-longitudinal, however, is much simpler with a polarization
vector which looks like a smooth deformation of the field of normal
vectors on the sphere (a hedgehog), see \refFig{quasiPressure}.

\section{The Green's function}
\label{gfSect}

We discuss a result based on the Radon transform. Numerically it is
convenient to express the Green's function as a certain integral over
the unit sphere. For a theoretical interpretation this integral is
transformed to a similar one  over the slowness surface. Using the Radon
transform the result in the frequency domain is
\rf{WangAchenbach,burridge}
\beq
\gf_+(\mathbf{x},\omega) =\gf^R_+(\mathbf{x},\omega)+\gf^S(\mathbf{x},\omega)
\eeq
with superscripts $R$ and $S$ indicating {\it regular \rm and \it
singular} part and subscript $+$ for {\it causal}. Here we have
changed dimensions compared to Refs.~\rf{WangAchenbach,burridge} by
solving for the Green's function for Eq.~\refeq{waveFreq} where $\rho$
has been divided out.  For convenience a derivation is discussed in
App.~\ref{radon}. The regular part is an integral over plane waves:
\be{radonGFReg1}
\gf^R_+(\mathbf{x},\omega) = \frac{i}{8 \pi^2} \int_{\mathbf n \in S^2} \sum_\alpha \frac{k_\alpha \proj^\alpha}{2  c_\alpha^2} \, e^{i k_\alpha |\mathbf{ n \cdot x}|} d\Omega \, .
\ee
Here $k_\alpha, c_\alpha$ and $\proj^\alpha$ refer to wave number,
phase velocity respective polarization projector for polarization
number $\alpha$ for the plane wave in question. The singular part
\be{radonGFSing}
\mathbf G^S(\mathbf{x},\omega) = \frac{1}{8 \pi^2} 
 \int_{\mathbf n \in S^2} \mathbf \Gamma^{-1} (\mathbf n) \, 
 \delta(\mathbf{ n \cdot x}) d\Omega
\ee
is purely real and corresponds to the {\it static} part of the Green's
function. The matrix $\Gamma$ is defined by
\begin{equation}
\Gamma_{ij} = \frac{c_{iklj}}{\rho} n_k n_l 
\label{g1}
\end{equation}
for which
\begin{equation}
\mathbf \Gamma(\mathbf n) = \sum_\alpha  c_\alpha^2 \, \proj^\alpha \, . 
\label{g2}
\end{equation}
We notice that \refeq{g1} follows from \refeq{dispRelation} with $\mathbf{k} =
\mathbf{n} \, k$ and $c = \omega/k$. 

The integrals above can be formulated in a more intrinsic form. Thus
instead of integrating over the unit sphere $S^2$ the dispersion
surface can be applied. Using {\it slowness}, Eq.~\refeq{radonGFReg1}
becomes,
\be{radonGFReg2}
\gf^R_+(\mathbf{x},\omega) = \frac{i}{8 \pi^2} \, \omega \, \int_{\mathbf n \in S^2} 
\sum_\alpha \frac{s_\alpha^3 \proj^\alpha}{2 } \, 
e^{i \omega  |\mathbf{s_\alpha  \cdot x}|} d\Omega\, .
\ee
Since the wave number and the angular velocity occur homogeneously in
the dispersion relation the corresponding slowness surface is
used. Applying polar reciprocity \refeq{polarRec}, $1 = \mathbf s \cdot
\mathbf v_g$ and denoting $\theta$ the angle between $\mathbf{s_\alpha}$
and $\mathbf{v_{g,\alpha}}$, the element of area $d\sigma_\alpha$ on this
surface is
\beq
d\Omega = \frac{\cos \theta}{s^2} d\sigma_\alpha =
\frac{d\sigma_\alpha}{\mathrm{v_{g,\alpha}} \, s_\alpha^3}
\eeq
so
\be{radonGFReg3}
\gf^R_+(\mathbf{x},\omega) = 
 \frac{i}{8 \pi^2} \, \omega \,  \sum_\alpha \int_{\mathbf s \in \Sigma_\alpha} 
 \frac{ \proj^\alpha}{2  \mathrm{v_{g,\alpha}}} \, 
 e^{i \omega  |\mathbf{s_\alpha  \cdot x}|} \, d\sigma_\alpha.
\ee
This can be written in a causal form as an integral over forward
pointing wave vectors
\be{radonGFReg4}
\gf^R_+(\mathbf{x},\omega) = \frac{i}{8 \pi^2} \, \omega \,  
\sum_\alpha \int_{\mathbf s \in \Sigma_\alpha^+} 
\frac{ \proj^\alpha}{\mathrm{v_{g,\alpha}}} \, 
e^{i \omega  \mathbf{s_\alpha  \cdot x}} \, d\sigma_\alpha \, .
\ee
The general result including the anti-causal Green's function reads
\be{g6}
\gf_\pm^R(x,y) = \pm i \, \omega \, \frac{1}{(2 \pi)^2} 
\sum_\alpha  \int_{\mathbf s \in \Sigma_\alpha^\pm} d\mathbf{\sigma}_\alpha \,
\mathbf{\frac{\proj_\alpha}{\mathrm 2 \, \mathrm{  v_{g,\alpha}} }} \,  
e^{i \omega \, \mathbf{s \cdot r}} \, ,
\ee
where $\Sigma_\alpha^\pm$ is the positive or negative of the surface
$\Sigma_\alpha$, respectively.  For the causal Green's function this
is to be understood over the upper hemisphere ($s_\| >0$). For the
anti-causal Green's function, the lower hemisphere, is the domain of
integration. Furthermore the group velocity is opposite of the
direction of observation, leading to a sign change of the
integrand. In conclusion, the {\it discontinuity} of the Green's function
\be{g7}
\Delta\gf(x,y) = \gf_+(x,y) - \gf_-(x,y) =  i \, \omega \, 
 \frac{1}{(2 \pi)^2} \sum_\alpha  \int_{\mathbf s \in \Sigma_\alpha} 
 d\mathbf{\sigma}_\alpha \, 
 \mathbf{\frac{e_\alpha}{\mathrm 2 \, \mathrm{  v_{g,\alpha}} }} \, 
 e^{i \omega \, \mathbf{s \cdot (x-y)}} \, 
\ee
is expressed as an integral over the whole slowness surface. Note here
the cancellation of the (real) singular part in the discontinuity of
the Green's function.

\section{Accumulation of data}
\label{data}

\FIG{
\rotatebox{0}{\hspace{3cm}\includegraphics[width=0.5\textwidth]{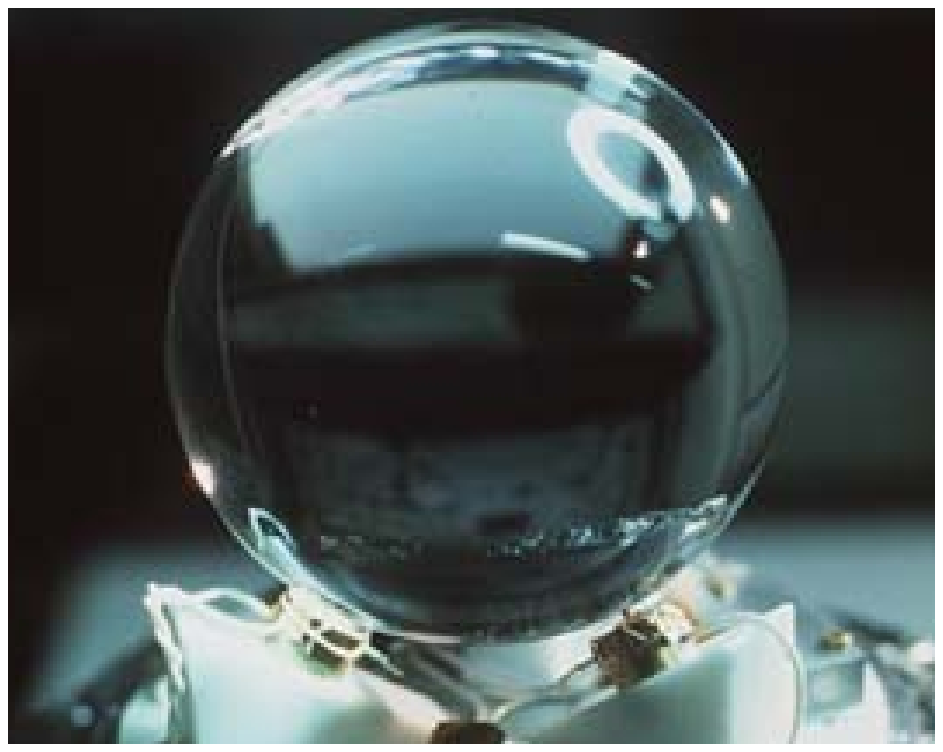}}
}{qsphere}{ Quartz sphere with supporting transducers.}{qsphere} 

In the experiments~\cite{MCK}, several shapes were studied. This
article concerns the spectrum of a particular quartz sphere. We shall
briefly sketch the experiments, further details will be given in
Ref.~\cite{MCK}.  The quartz sphere is supported by three transducers,
see Fig.~\ref{qsphere}. The transducers excite the sphere via
piezo-electric coupling. The data obtained in the experiment is an
arbitrary constant times the complex ratio of the outgoing to the
incoming {\it electric} signal.  As the data are complex the actual
phase of the response is available. The setup and the measurement
techniques are extensions of previous work~\rf{quartz}. The resolution
of the setup, characterized by the $Q$-value, is very high. A typical
resonance measured in experiment is shown in
Fig.~\ref{resonanceFig}. Due to the high $Q$-value {\it ringing} is
observed because the transducer frequency changes at a constant
rate. The actual position of the {\it single} resonance is to the left
of the observed peaks as the frequency is swept upwards.  Since
occasionally two or perhaps even more resonances are close together,
the observed signal appears as superpositions of such. The resonances
are fitted by using a recognition code~\cite{MO} which is based on
maximum entropy principles. The code fits for each resonance the
position in frequency, the decay width and its (complex) amplitude. It
is capable of treating individual resonances with ringing and of
superimposing them if they are overlapping. Good performance is
observed for the real and imaginary part of the response.

The quartz sphere was manufactured with a high precision.  The
spherical shape ensures the discrete symmetry of the crystal, which is
described by the symmetry group $D_3$. However, $D_3$ is broken to the
symmetry group $C_3$ due to the positions of the transducers in the
experiment. Moreover, despite the high precision manufacturing, the
shape cannot be perfectly spherical. Its diameter has a relative
precision of $5 \cdot 10^{-5}$. At high frequencies this should lead
to observable splittings. Finally, gravity leads to an oblate
deformation of the sphere. These effects further break the
symmetry. Hence, the vast majority, if not all, of the symmetry
degeneracies are expected to be broken.

\FIG{
\rotatebox{0}{\includegraphics[width=0.9\textwidth]{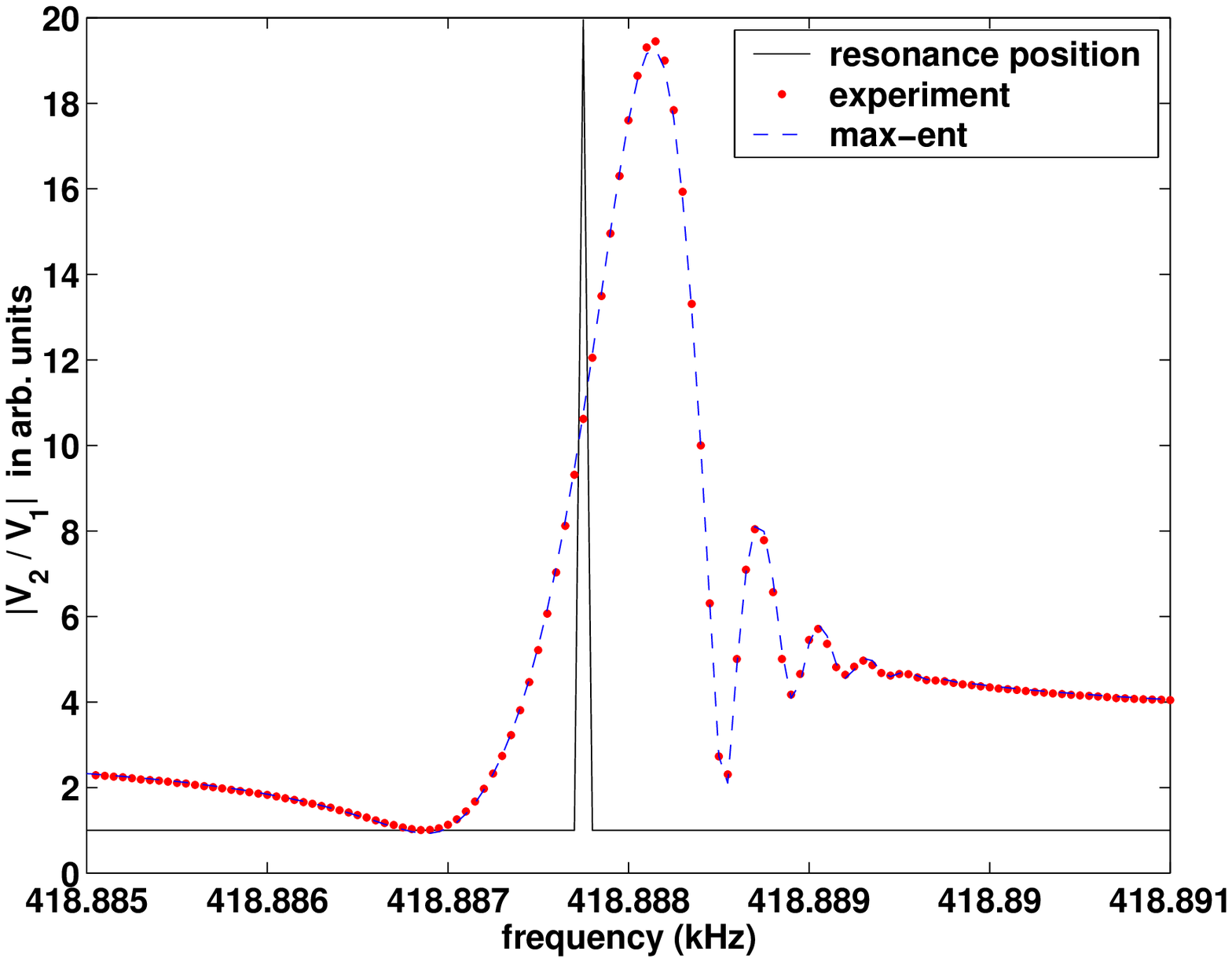}}
}{resonanceFig}{ Resonance with ringing due to finite sweep velocity
and high $Q$-value}{resonanceFig}

\section{Weyl contribution}
\label{weylSect}

We now discuss the counting function. We work out the leading
approximation in Sect.~\ref{sec51} and evaluate it numerically in
Sect.~\ref{sec52}. In Sect.~\ref{sec52}, we compare to data obtained
in the measurements mentioned in the previous Sect.~\ref{data}.

\subsection{Approximation to the counting function}
\label{sec51}

The smooth or Weyl part of the level density is the easiest to access. The
fluctuations play an important role as well. In fact, the oscillating
part of the level density  becomes of the same order as the smooth part.  Here,
however, the focus is on the smooth part.  The level density  is
written formally as
\beq
d(\omega) = \sum_i \delta(\omega -\omega_i) \, .
\eeq
This can be found from the discontinuity of the Green's function as:
\beq
d(\omega^2) = \frac{1}{2 \pi i} \, \Tr \, \Delta\gf(\omega^2)
\eeq
since we here use a Green's function with respect to the squared
angular velocity. Now the Green's function is approximated with the
corresponding free function. This step is discussed in \cite{Balian}
and below in this section and corresponds to that in the interior
parts of the resonator the propagation is roughly free. This gives
\beq
d(\omega^2) =   V_x \,  \omega \, \frac{1}{(2 \pi)^3} \sum_\alpha  
               \int_{\mathbf s \in \Sigma_\alpha} d\mathbf{\sigma}_\alpha \,
               \mathrm{\frac{1}{\mathrm 2 \, \mathrm{  v_{g,\alpha}} }}  \, . 
\eeq
The volume of the resonator is denoted  $V_x$. The counting function
\beq
N(\omega)=N(\omega^2) =   \int_0^{\omega^2} d(\omega'^2) \,d\omega'^2
\eeq
becomes
\beq
N(\omega) = V_x \, \frac{\omega^3}{3} \, 
            \frac{1}{(2 \pi)^3} \sum_\alpha  
            \int_{\mathbf s \in \Sigma_\alpha} d\mathbf{\sigma}_\alpha
            \, \mathrm{\frac{1}{ \mathrm{  v_{g,\alpha}} }}  \, . 
\eeq
From Eq.~\refeq{polarRec} and geometrical considerations of cones the
volume element on a given slowness sheet is
\beq 
dV_\alpha = \frac{1}{3} \frac{d\mathbf{\sigma}_\alpha}{\mathrm{v_{g,\alpha}}}
\eeq
and therefore
\beq
N(\omega)=  V_x    \, \sum_\alpha V_\alpha  \, \frac{\omega^3}{(2 \pi)^3}  \,  
\eeq
in terms of the volumes of the slowness sheets, $V_\alpha$.  Somewhat
simpler, using frequency $f$
\beq
N(f)=  V_x    \, \sum_\alpha V_\alpha  \, {f^3}  \, . 
\label{weylVolTerm}
\eeq
Thus we have a re-derivation of the well-known result: namely, the
number of states equals to leading order the available phase space
volume. 

Although we deal with the anisotropic case, it is instructive to see
that the formulae reduce to the result for the isotropic case,
\beq
N(f)=  \frac{4 \pi}{3} \,  
       V_x \left( \frac{1}{c_L^3}+\frac{2}{c_T^3} \right)  \, {f^3}  \, ,
\eeq
where $c_L$ and $c_T$ are the longitudinal and the transverse
sound velocities, respectively.
 
In solid state physics the procedure is different. The resonators are
unit cells in crystal lattices. This implies a periodicity assumption
of the crystal in contrast to our case of a finite system. Lattice
vibrations are connected to the concept of heat and the counting
function for vibrations leads to the famous Debye law for the heat
capacity. This is derived in Ref.~\cite{AM}.

We will now discuss heuristically why the above derivation is nothing
but the leading approximation and correponds to the smooth part of the
counting function. Since the trace of the Green's function is taken,
it is mainly the local propagation between nearby points which is
probed. However, also the possibility of propagation via reflections
from the boundary remains.  Contributions here come from periodic
orbits $p$ but are typically of lesser magnitude than the direct zero
length contribution.  Ultimately such terms lead to oscillatory
contributions to the spectral density. Thus expanding around periodic
orbits using the method of stationary phase leads to terms of the form
of amplitudes $A_p$ times phases
\beq
A_p \, \exp(i \omega T_p+\dots) \, ,
\eeq
where the (dimensionless) classical action $k \, L_p = \omega T_p $ in
terms of the length/period of the orbit controls the phase of the
fluctuations. Averaged over a sufficently large frequency interval the
fluctuations dissappear and what remains is the part from zero length
orbits. The latter do not fluctuate and represents the average
contribution.

\subsection{Numerical calculation and prediction}
\label{sec52}

At room temperature $T_r = 22^\circ\mathrm{ C}$, the mass density is
$\rho = 2.6485 \, \mathrm{g/cm^3}$, and the diameter of the sphere
measures $87.75 \,\, \mathrm{mm}$. The elastic constants for quartz at
room temperature are listed in \refTab{eltensorTab},
\begin{table}
\caption{Elastic constants for quartz at room temperature.}
\begin{center}
\begin{tabular}{|c|c|c|c|c|c|c|}
\hline
units & C11 & C12 & C13 & C14 & C33 & C44 \\
\hline
$10^{11} \, \mathrm{N/m}^2 $ & 0.868  & 0.0704 & 0.1191 & -0.1804 & 1.0575 & 0.5820 \\
\hline
\end{tabular}
\end{center}
\label{eltensorTab}
\end{table}
They are given in Voigt notation. Pairs of Cartesian
indices $ij$ with $i,j \in \{ x,y,z \}$ are numbered as follows $xx =
1$, $yy = 2$, $zz = 3$, $yz = 4$, $zx = 5$ and $xy = 6$. Thus for
example $c_{xxyz}$ is denoted C14. 

We now calculate the leading Weyl term \refeq{weylVolTerm}.  The
volume in slowness space is conveniently expressed as an integral over
the unit sphere,
\beq
\label{volslow}
V_s = \sum_\alpha V_\alpha=\sum_\alpha \frac{1}{3} \int_{\mathbf n \in S^2} s_\alpha^3 \, d\mathbf n \, .
\eeq
The three lengths of the slowness vectors are found from
Eqs.~\refeqs{slowmat}{slowsurf}. In particular, the squares of the
velocities are found from Eqs.~(\ref{g1}) and~(\ref{g2}). Therefore
Eq.~\refeq{volslow} reduces to
\beq
V_s = \frac{1}{3} \int_{\mathbf n \in S^2} 
      \tr{\left( \mathbf \Gamma(\mathbf n)^{-3/2} \right)} \, d\mathbf n \, ,
\eeq
which is easily calculated from the eigenvalues of $\mathbf
\Gamma(\mathbf n)$. Finally the symmetry can be exploited: In the case
of quartz it is enough to integrate over a fundamental region $0 \le
\phi \le 2 \pi/3$ and $0 \le \theta \le \pi/2$ and multiply by six. Using
the elastic constants in Tab.~\ref{eltensorTab}, we find the volume in slowness space  by numerical
integration,
\beq
V_s = 1.4909 \cdot 10^{-10} \, \mathrm{(sec/m)}^3\, .
\eeq
This volume may be used for various shapes of resonators upon
multiplication with the particular volumes in physical
$x$-space. Hence, combining everything, the number of states for our
sphere should grow as
\beq
\label{cubefit}
N(f) = A \, f^3 + {\cal O}(f^2)\, 
\eeq
with the constant 
\begin{equation}
A = 5.274 \cdot 10^{-14} \, \mathrm{sec}^3  
\label{atheo}
\end{equation}
as the result of our computation.

The elastic constants change with temperature. In App.~\ref{appB},
they are given in terms of an expansion. As the experiment was
performed at $T = 30^\circ\mathrm{ C}$, we also calculate the volumes
in slowness space with these corrected constants. However, it turns
out that the relative difference to the calculation for room
temperature is of the order of $10^{-8}$, that is, within the
numerical errors. Similarly, the temperature dependence of the mass
density, see App.~\ref{appB}, has no influence either.

\subsection{Comparing with experiment}
\label{sec53}

\FIG{
\rotatebox{0}{\includegraphics[width=0.8\textwidth]{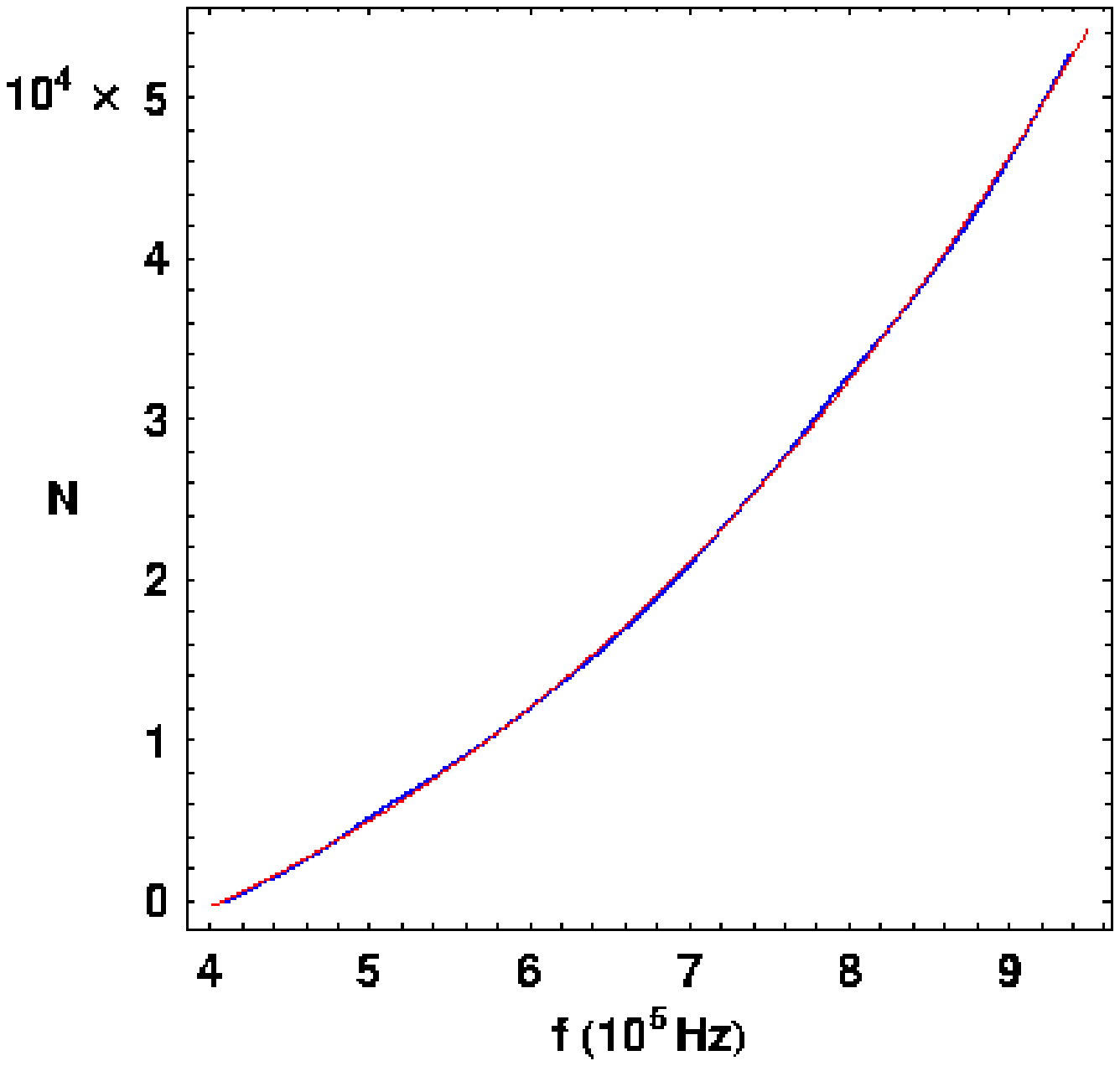}}
}{stairExpFit}{ Spectral staircase: experiment and fit}{stairExpFit}

The spectrum used here was measured from 410 kHz to 960 kHz. The
resonances were identified by means of the recognition
routine~\cite{MO} mentioned in Sec.~\ref{data}.  Roughly 52000
resonances were found.  Here the bulk term in Eq.~\refeq{cubefit}
contributes with at least 43000 resonances.  The staircase for the
experiment is fitted with a cubic polynomial, see \refFig{stairExpFit}.  For the leading
coefficient we find
\begin{equation}
A_{\rm fit} = 5.030 \cdot 10^{-14} \, \mathrm{sec}^3 \ .
\end{equation}
This is in good agreement with the theoretical result~(\ref{atheo}).
The relative error is $4.6\% $. 

The difference between the fitted and the experimental counting
function is plotted in \refFig{stairFluct}. The sizable deviation
between fit and experiment is striking.  It resembles a sawtooth with
very rapid fluctuations on top. We Fourier transform this oscillating
structure and obtain the period spectrum, see \refFig{stairPowerSpec}.
From this we estimate the time corresponding to the largest peak to
approximately $2$ $\mu \mathrm{sec}$. Also other peaks at $4, 9, 13,
17, 20, \dots \mu \mathrm{sec}$ are observed. As the frequency
interval sampled is approximately $500$ $\mathrm{KHz}$ the resolution
is roughly $2$ $\mu \mathrm{sec}$.  It is tempting to interpret the
largest peak as due to the shortest bouncing-ball-type-of mode, that
is, a wave packet bouncing back and forth along some diameter of the
sphere.  If we take the velocities to lie between $5$
$\mathrm{km/sec}$ to $10$ $\mathrm{km/sec}$ the shortest period is
around $17.8$ $\mu \mathrm{sec}$. Therefore the peak corresponding to
the sawtooth oscillations is not likely to be due to such a mode and
remains unexplained so far. Likewise the remaining short times less
than the estimate $17.8$ $\mu \mathrm{sec}$ are not explained.

\FIG{
\rotatebox{0}{\includegraphics[width=0.9\textwidth]{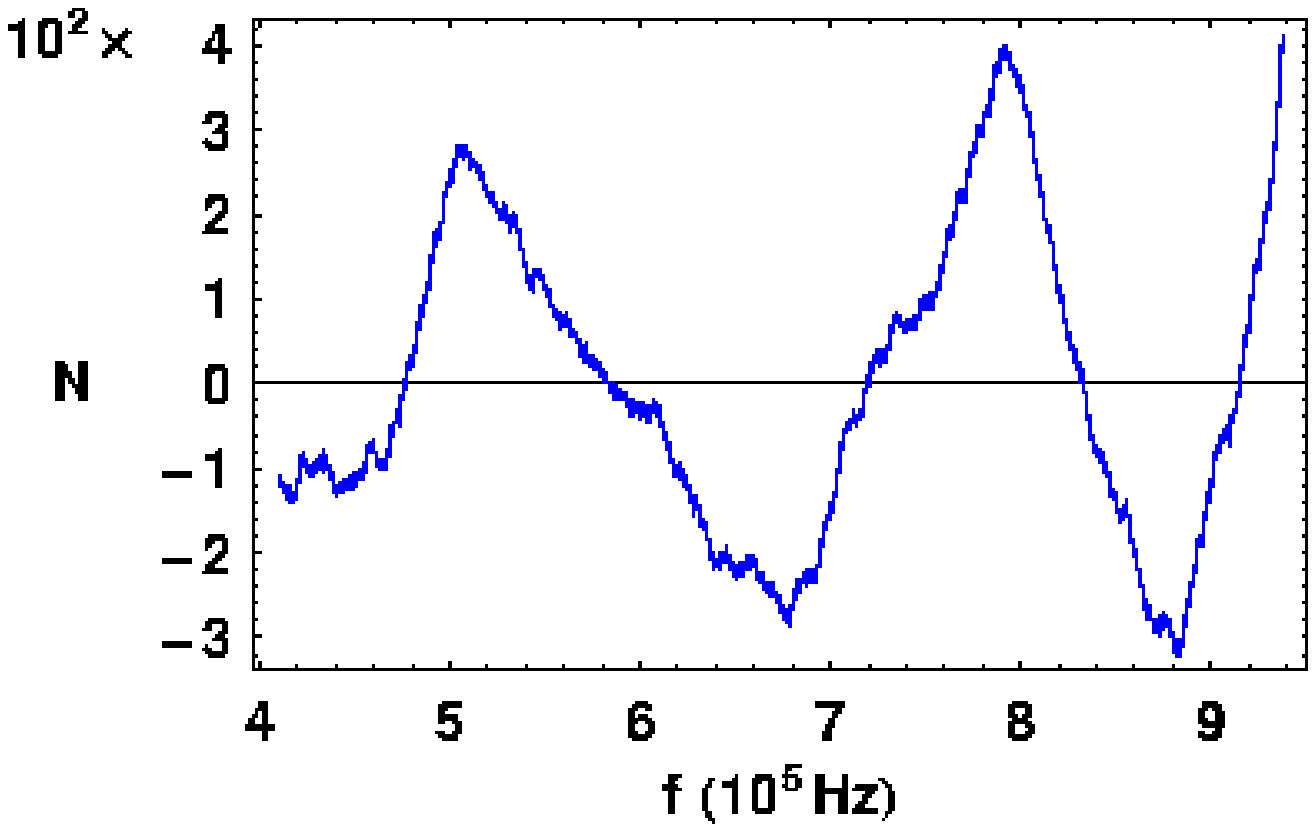}}
}{stairFluct}{ Fluctuations of the counting function}{stairFluct}

\FIG{
\rotatebox{0}{\includegraphics[width=0.9\textwidth]{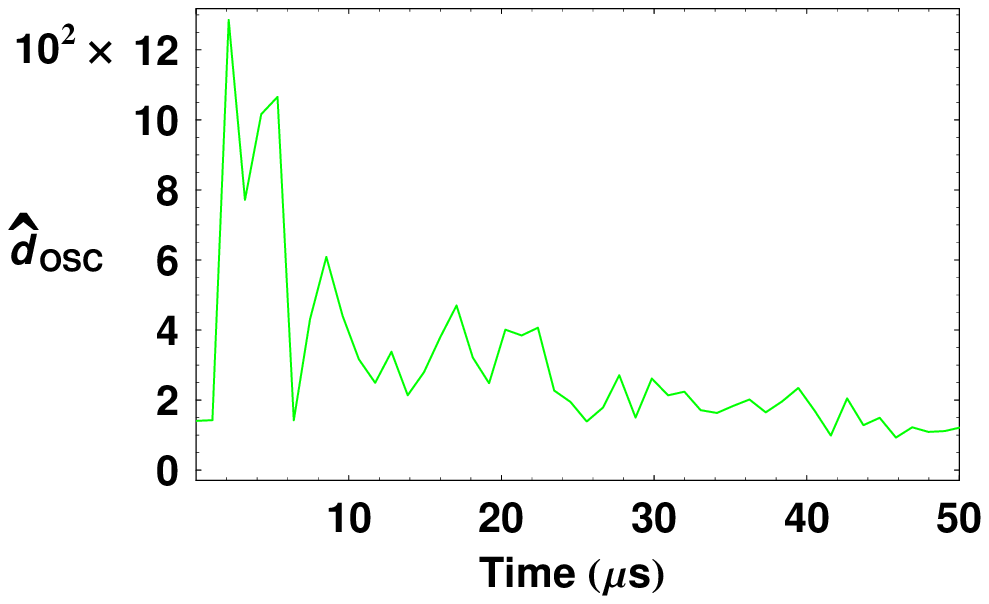}}
}{stairPowerSpec}{ Period Spectrum}{stairPowerSpec}

\section{Summary and conclusion}
\label{summarySect}

We have discussed the experimental spectrum of a quartz sphere from
the point of view of the spectral counting function. The classical
phase space volume gives indeed  the leading behavior of this
object. On calculating this numerically we find it in   good agreement with
the experimentally fitted term. This paves the way for
further studies of elastic spectra, such as boundary corrections to
the counting function\rf{Yale,SAFAROV} and periodic orbit theory for
the fluctuations in the level density\rf{couch,disc,HB}.

Due to the boundary  further resonances  should be present. Less is
not expected as the boundary conditions allow for surface waves in
analogy with Neumann conditions for the Helmholtz case.  It is
currently an open problem to incorporate this surface contribution in
the anisotropic case. Nevertheless such surface modes couple to the
bulk and are not exponentially damped in this elastic case. Therefore
experiments should access these states and thereby the surface
correction to the leading volume term as well.

The fluctuating part would involve recent semiclassical methods. Here
individual closed rays, periodic orbits, control the oscillations in
the level density. In particular this has been shown not just for
quantum mechanical spectra but also more recent in two dimensional
elastodynamic resonators. The step to three dimensions plus anisotropy
has yet to be made.

\section*{Acknowledgments}

N.S. and T.G. acknowledge discussions with S. Creagh and
J.D. Achenbach and financial support from the Rosenfeld foundation,
the Crafoord Foundation and from Det Svenska Vetenskapsr\aa det,
respectively.  M.O. thanks the Division of Mathematical Physics, LTH,
for hospitality during his visits to Lund.

\begin{appendix}

\section{Radon transform and the Green's function}
\label{radon}

Following~\refRef{WangAchenbach} we first solve the Radon
transformed time problem. Next the solution is transformed back and
finally Fourier transforming gives the frequency dependent Green's
function.

\subsection{Radon transform}

The equation $s={\bf n \cdot x}$ describes a plane with the normal
vector ${\bf n}$ in the space of the position vector ${\bf x}$.  The
plane cuts the direction defined by ${\bf n}$ in the height $s$. The
Radon transform $\hat{f}(s,{\bf n})$ of a function $f({\bf x})$ is the average
\beq
{\cal R} f(s,{\bf n}) = \hat{f}(s,{\bf n}) = \int f({\bf x}) 
\delta(s - {\bf n \cdot x}) \, d{\bf x}
\eeq
over the plane for a given height $s$.  The Radon transform of the
spatial derivatives of the function $f({\bf x})$ fulfill
\beq
{\cal R} \left(\frac{\partial f}{\partial x^i}\right)(s,{\bf n})= n_i
\, (\partial_s {\cal R} f)(s,{\bf n}) \ .
\eeq
We now apply this to the time dependent elastic
problem~(\ref{waveTime}). We denote the Green's function ${\mathbf
G}(\mathbf x,t)$ and its Radon transform $\hat{\mathbf G}(s,{\bf n})$.
We find
\beq
\left({\mathbf \Gamma}(\normal) \, \partial_s^2 - {\mathbf 1} \, 
 \partial_t^2 \right) \cdot \hat{\mathbf G} = - {\mathbf 1} \,
 \delta(s) \delta(t) \, ,
\eeq
where the matrix ${\mathbf \Gamma}$ is defined in Eqs.~(\ref{g1})
and~(\ref{g2}).  We observe that $\hat{\mathbf G}(s,{\bf n})$ also
depends on time $t$.  The initial condition is taken as causal such
that $\hat{\mathbf G}(s,{\bf n})=0$ for $t<0$.  Necessary and
sufficient conditions for solving this problem are found by projecting
onto the subspaces defined by the individual polarizations:
\beq
\left(c_\alpha^2 \, \partial_s^2 -  \partial_t^2 \right) 
 \proj^\alpha \cdot \hat{\mathbf G}(s,{\bf n}) = -\proj^\alpha \,
 \delta(s) \delta(t) \, ,
\eeq
where $c_\alpha$ is the phase velocity associated with polarization
$\alpha$ and $\proj^\alpha$ is the corresponding projector.  Invoking
d'Alembert's solution
\beq
\proj^\alpha \cdot \hat{\mathbf G}(s,{\bf n}) = 
\proj^\alpha \, \frac{H(t)}{2 c_\alpha}\left(H(s+c_\alpha t) -H(s-c_\alpha t)\right)
\eeq
with $H$ the Heaviside unit step function, we find
\be{dAlembProj}
\hat{\mathbf G}(s,{\bf n}) = \sum_\alpha \proj^\alpha \, \frac{H(t)}{2 c_\alpha} 
\left(H(s+c_\alpha t) -H(s-c_\alpha t)\right) \, ,
\ee
which is the full solution.

\subsection{Inverse Radon transform}

The inverse Radon transform is given by 
\be{invRad}
f(\mathbf{x}) = -\frac{1}{8 \pi^2}\int_{\mathbf n \in S^2} d\Omega \, \, 
 \frac{d^2({\cal R} f)(s,{\bf n})}{ds^2}|_{s ={\bf n \cdot x}} \, 
\ee
following from the identity
\beq
\delta(\mathbf{x}) = -\frac{1}{8 \pi^2}\int_{\mathbf n \in S^2} d\Omega \, \, 
\delta''(\mathbf{n \cdot x}) \, . \eeq
The latter is proved using
\[
2 \pi |\mathbf{x}| = \int_{\mathbf n \in S^2 } d\Omega \, \, |\mathbf{n \cdot x}| \, 
\]
and the identities $|x|'' = 2 \delta(x)$, $\Delta
\frac{1}{|\mathbf{x}|} = - 4 \pi \delta(\mathbf{x})$ and $\Delta
|\mathbf{x}| = \frac{2}{|\mathbf{x}|}$. Hence, we find the Green's
function in configuration space,
\bea
{\mathbf G}(\mathbf x,t) &=& -\frac{H(t)}{16 \pi^2} \int_{\mathbf n \in S^2} d\Omega \,   \sum_\alpha \frac{\proj^\alpha}{c_\alpha^2} \, \partial_t(\delta(c_\alpha t +\mathbf{n \cdot x}) +  \delta(c_\alpha t -\mathbf{n \cdot x}) ) \continue 
                         &=& -\frac{H(t)}{8 \pi^2} \int_{\mathbf n \in S^2} d\Omega \,   \sum_\alpha \frac{\proj^\alpha}{c_\alpha^2} \, \partial_t \delta(c_\alpha t -\mathbf{n \cdot x}) \, ,
\label{gTime}
\eea
by inverting Eq.~(\ref{dAlembProj}).

\subsection{Frequency domain}

The frequency dependent Green's function $\gf(x,\omega)$ is obtained by fourier transforming \refeq{gTime} with respect to time:
\beq
{\mathbf G}(\mathbf x,\omega) = -\frac{1}{8 \pi^2} \int_{\mathbf n \in S^2} d\Omega \,   \sum_\alpha \frac{\proj^\alpha}{c_\alpha^2} \, \int_{-\infty}^{\infty} dt \, e^{i \omega t}  H(t) \, \partial_t \delta(c_\alpha t -\mathbf{n \cdot x})   \, .                       
\eeq
The inner integral is found by partial integration: the time
derivative will act either on the exponential or the step function
leading to the regular respective singular contribution 
\bea
\gf^R_+(\mathbf{x},\omega) &=& \frac{i}{8 \pi^2} \int_{\mathbf{ n \cdot x} >0} \sum_\alpha \frac{k_\alpha \proj^\alpha}{  c_\alpha^2} \, e^{i k_\alpha \mathbf{ n \cdot x}} d\Omega \continue
                         &=& \frac{i}{8 \pi^2} \int_{\mathbf n \in S^2} \sum_\alpha \frac{k_\alpha \proj^\alpha}{2  c_\alpha^2} \, e^{i k_\alpha |\mathbf{ n \cdot x}|} d\Omega \, 
\eea
and 
\[
\mathbf G^S(\mathbf{x},\omega) = \frac{1}{8 \pi^2} \int_{\mathbf n \in S^2} \mathbf \Gamma^{-1} (\mathbf n) \, \delta(\mathbf{ n \cdot x}) d\Omega \, .
\]
The anti-causal Green's function is found similarly by reversing the
time in say Eq.~\refeq{dAlembProj}. In particular the same singular
part is otained whereas the regular part is the complex conjugate of
the causal regular part.

\section{Temperature dependence}
\label{appB}

The resonator is kept at fixed temperature 
\beq
T = 30^\circ\mathrm  C
\eeq
with a precision on the order of a milli-Kelvin. This is slightly
higher than the room temperature $T_r=22^\circ \mathrm C$, at which
the quartz sphere was manufactured.  This causes a change in the
elastic constants described below in terms of a power series expansion
around room temperature,
\beq
c_{ijkl}(T) = c_{ijkl,0} + c_{ijkl,1}(T-T_r) + c_{ijkl,2}(T-T_r)^2 
                         + c_{ijkl,3}(T-T_r)^3 + \ldots
\eeq
We took the coefficients of this expansion from Ref.~\rf{bechmann},
they are listed in Tab.~\ref{tabb1},
\begin{table}
\caption{Coefficients of temperature expansion for elastic constants
         of quartz.}
\begin{center}
\begin{tabular}{|c|c|c|c|c|c|c|c|}
\hline
order & units & C11 & C12 & C13 & C14 & C33 & C44 \\
\hline
0 & $10^{11} \,  N/m^2$ & 0.868  & 0.0704 & 0.1191 & -0.1804 & 1.0575 & 0.5820 \\
1 & $\times 10^{-6}/K$  & -48.5 & -3000 & -550 & 101 & -160 & -177 \\
2 & $\times 10^{-9}/K^2 $ & -107 & -3050 & -1150 & -48 & -275 & -216 \\
3 & $\times 10^{-12}/K^3 $ & -70 & -1260 & -750 & -590 & -250 & -216 \\  
\hline 
\end{tabular}
\end{center}
\label{tabb1}
\end{table}
Furthermore a change in mass density takes place. In Tab.~\ref{tabb2}, we
list the coefficients of an expansion around room temperature,
\begin{table}
\caption{Coefficients for the temperature expansion of the
         mass density of quartz.}
\begin{center}
\begin{tabular}{|c|c|c|}
\hline
order & units & $\rho$  \\
\hline
0 & $10^{3} \, kg/m^3 $ & 2.6485   \\
1 & $\times 10^{-6}/K$  & -34.80 \\
2 & $\times 10^{-9}/K^2 $ & -30.04 \\
3 & $\times 10^{-12}/K^3 $ & 49.08  \\  
\hline 
\end{tabular}
\end{center}
\label{tabb2}
\end{table}
Finally, there is also a shape deformation, expressed in terms of the tensor 
of thermal expansion according to $u'_i = \alpha_{i j} u_j$.

The effect of temperature change is therefore mainly to change the
shape and volume of the $k$-part of the available phase space volume,
namely the slowness surface. Thus the first table is most important
for our purposes. We have calculated the new volume in slowness space
at the corresponding experimental temperature to cubic order in the
temperature change. The other, relative changes turn out to be
negligible, namely of the order $10^{-8}$.

\end{appendix}


\begin{thebibliography}{999}

\bibitem{AM}          N.W. Ashcroft and N.D. Mermin, 
                      {\em Solid State Physics} (Holt, Rinehart and
                      Winston, New York 1976).
\bibitem{Predrag}     P. Cvitanovi\'c, R. Artuso, R. Mainieri 
                      and G. Vattay, 
                      {\it Classical and Quantum Chaos}, 
                      {\tt www.nbi.dk/ChaosBook/}, 
                      Niels Bohr Institute, Copenhagen, 2002.
\bibitem{gutbook}     M.C. Gutzwiller, 
                      {\em Chaos in Classical and Quantum Mechanics} 
                      (Springer, New York 1990).
\bibitem{brack}       M.~ Brack and R.K. Bhaduri, 
                      {\it Semiclassical Physics} 
                      (Addison-Wesley, Reading 1997).
\bibitem{Stock}       H.-J. St\"ockmann, 
                      {\em Quantum Chaos: An Introduction} 
                      (Cambridge University Press, 1999).
\bibitem{BaltHil}     H.P.~Baltes and E.R.~Hilf, 
                      {\em Specta of finite systems} 
                      (Bibliographisches Institut, Mannheim 1976).
\bibitem{GeoQu}       F. Faure and B. Zhilinskii,  
                      Phys. Lett. {\bf A302}, 242 (2002). 
\bibitem{Weaver}      R.L. Weaver,
                      J. Acoust. Soc. Am. {\bf 85}, 1005 (1989).
\bibitem{alu}         C. Ellegaard, T. Guhr, K. Lindemann, 
                      H.Q. Lorensen, J. Nyg\aa rd and M. Oxborrow,
                      Phys. Rev. Lett. {\bf 75}, 1546  (1995).
\bibitem{quartz}      C. Ellegaard, T. Guhr, K. Lindemann, 
                      J. Nyg\aa rd and M. Oxborrow, 
                      Phys. Rev. Lett. {\bf 77}, 4918 (1996).
\bibitem{para}        P. Bertelsen, C. Ellegaard, T. Guhr, 
                      M. Oxborrow and K. Schaadt,
                      Phys. Rev. Lett. {\bf 83}, 2171 (1999).
\bibitem{waves}       K. Schaadt, T. Guhr, C. Ellegaard and 
                      M. Oxborrow,
                      Phys. Rev. {\bf E68}, 036205 (2003). 
\bibitem{couch}       L. Couchman, E. Ott and T.M. Antonsen, Jr.,
                      Phys. Rev. {\bf A46}, 6193 (1992).
\bibitem{disc}        N. S\o ndergaard and G. Tanner, 
                      Phys. Rev. {\bf E66}, 066211 (2002).
\bibitem{HB}          E. Bogomolny and E. Hugues,
                      Phys. Rev. {\bf E57} 5404 (1998).
\bibitem{MCK}         M.~Oxborrow, C.~Ellegaard and K.~Schaadt, 
                      in preparation.
\bibitem{MEMS}        A. N.~Cleland,{\em Foundations of Nanomechanics}
                      (Springer, Berlin 2002).
\bibitem{lAndl}       L.D. Landau and E.M. Lifshitz, 
                      {\it Theory of Elasticity} 
                      (Pergamon, Oxford 1959).
\bibitem{auld}        B.A. Auld, {\em Acoustic Fields and Waves in Solids I,II}                       (John Wiley and Sons, New York 1973).
\bibitem{musgrave}    M.J.P. Musgrave, {\em Crystal Acoustics}
                      (Holden-Day, San Fransisco 1970).
\bibitem{WangAchenbach} C.Y. Wang, A. Saez and J.D. Achenbach,    
                      {\em 3-d elastodynamic Green's functions for BEM applications 
                      to anisotropic solids}, 
                      IUTAM Symposium on Anisotropy, Inhomogeneity and Nonlinearity 
                      in Solid Mechanics, Eds. D.F. Parker and A.H. England, 
                      307 (1995).
\bibitem{burridge}    R. Burridge,  
                      {Quart. Journ. Mech. and Appl. Math \bf 20}, 41 (1967).
\bibitem{MO}          M.~Oxborrow, 
                      to be published.
\bibitem{Balian}      R. Balian and C. Bloch,
                      Ann. Phys. {\bf 60}, 401 (1970).  
\bibitem{Yale}        M. Dupuis, R. Mazo and L. Onsager, 
                      Journ. of Chem. Phys. {\bf 33}, 1452 (1960).
\bibitem{SAFAROV}     Y. Safarov and D.~Vassiliev,   
                      in {\em Spectral Theory of Operators}, 
                      edited by S.Gindikin, 
                      AMS Translations,  ser.2, {\bf 150} (1992).
\bibitem{bechmann}    R.~Bechmann, 
                      Phys. Rev. {\bf 110}, 1060 (1958).
\end{thebibliography}
\end{document}